\documentclass[aps,prl,twocolumn,amsmath,amssymb,amsfonts,superscriptaddress,floatfix,showpacs]{revtex4-1}
\usepackage{graphicx}% Include figure files  
\usepackage{dcolumn}% Align table columns on decimal point  
\usepackage{bm}% bold math  
\usepackage{rotating} % rotate figure  
\usepackage{epsfig}
\usepackage{diagbox}
\usepackage[version=3]{mhchem} % Formula subscripts using \ce{}
\usepackage{cancel}
\usepackage{caption}
\usepackage{multirow}

\usepackage[normalem]{ulem}
\usepackage{color}
\usepackage[dvipsnames]{xcolor}

\newcommand*{\LightComments}{}%

\newcommand{\hide}[1]{}

\ifdefined\NoComments

\newcommand{\comment}[1]{}
\else
\ifdefined\LightComments

\newcommand{\comment}[1]{\textcolor{gray}{\emph{#1}}}
\else

\newcommand{\comment}[1]{\textcolor{gray}{\emph{#1}}}
\fi
\fi
\definecolor{Mygrey}{gray}{0.80}
\newcommand{\vp}{\mathbf{p}}
\newcommand{\vq}{\mathbf{q}}

\newcommand{\Pt}{\tilde{P}}
\newcommand{\iexpect}[1]{\langle #1 \rangle}

\newcommand{\ads}{\text{ads}}
\newcommand{\PBE}{\text{PBE}}
\newcommand{\PBED}{\text{PBE+D2}}
\newcommand{\RPA}{\text{RPA}}

\RequirePackage[normalem]{ulem} %DIF PREAMBLE
\RequirePackage{color}\definecolor{MAG}{rgb}{1,0,1}\definecolor{TEAL}{rgb}{0,0.5,0.5} %DIF PREAMBLE
\begin{document}

%%%%%%%%%%%%%%%%%%%%%%%%%%%%%%%%%%%%%%%%%%%%%%%%%%%%%%%%%%%%%%%%
\title{Bridging molecular dynamics and correlated wave-function methods for accurate finite-temperature properties}
%%%%%%%%%%%%%%%%%%%%%%%%%%%%%%%%%%%%%%%%%%%%%%%%%%%%%%%%%%%%%%%%
\date{\today}
\author{Dario Rocca}
\email{dario.rocca@univ-lorraine.fr}
\affiliation{Universit\'{e} de Lorraine, LPCT, UMR 7019, 54506 Vand\oe{}uvre-l\`{e}s-Nancy, France}
\affiliation{CNRS, LPCT, UMR 7019, 54506 Vand\oe{}uvre-l\`{e}s-Nancy, France}
\author{Anant Dixit}
%\email{dixit1@univ-lorraine.fr}
\affiliation{Universit\'{e} de Lorraine, LPCT, UMR 7019, 54506 Vand\oe{}uvre-l\`{e}s-Nancy, France}
\affiliation{CNRS, LPCT, UMR 7019, 54506 Vand\oe{}uvre-l\`{e}s-Nancy, France}
\author{Michael Badawi}
%\email{michael.badawi@univ-lorraine.fr}
\affiliation{Universit\'{e} de Lorraine, LPCT, UMR 7019, 54506 Vand\oe{}uvre-l\`{e}s-Nancy, France}
\affiliation{CNRS, LPCT, UMR 7019, 54506 Vand\oe{}uvre-l\`{e}s-Nancy, France}
\author{S\'ebastien~Leb\`egue}
%\email{sebastien.lebegue@univ-lorraine.fr}
\affiliation{Universit\'{e} de Lorraine, LPCT, UMR 7019, 54506 Vand\oe{}uvre-l\`{e}s-Nancy, France}
\affiliation{CNRS, LPCT, UMR 7019, 54506 Vand\oe{}uvre-l\`{e}s-Nancy, France}
\author{Tim Gould}
%\email{t.gould@griffith.edu.au}
\affiliation{Qld Micro- and Nanotechnology Centre, %
Griffith University, Nathan, Qld 4111, Australia}
\author{Tom\'{a}\v{s} Bu\v{c}ko}
\email{bucko19@uniba.sk}
\affiliation{Department of Physical and Theoretical Chemistry, Faculty of Natural Sciences, Comenius University in Bratislava, Ilkovi\v{c}ova 6, SK-84215 Bratislava, Slovakia}
\affiliation{Institute of Inorganic Chemistry, Slovak Academy of Sciences, D\'{u}bravsk\'{a} cesta 9, SK-84236 Bratislava, Slovakia}
%%%%%%%%%%%%%%%%%%%%%%%%%%%%%%%

\begin{abstract}
  We introduce the ``selPT'' perturbative approach, based on ab initio
  molecular dynamics (AIMD), for computing accurate finite-temperature
  properties by efficiently using correlated wave-function methods.
  We demonstrate the power of the method by computing 
  prototypical molecular enthalpies of adsorption in zeolite
  (CH$_4$ and CO$_2$ on protonated chabazite at 300~K)
  using the random phase approximation. Results are in excellent
  agreement with experiment. The improved accuracy provided by selPT
  represents a crucial step towards the goal of truly quantitative
  AIMD prediction of experimental observables at finite temperature.
\end{abstract}

%\pacs{71.15.-m, 71.15.Mb, 71.20.Nr}
\pacs{}
%PACS, the Physics and Astronomy Classification Scheme.
%\preprint{APS/123-QED}

\maketitle

%%%%%%%%%%%%%%%%%%%%%%%%%%%%%
%\section{Introduction}
%%%%%%%%%%%%%%%%%%%%%%%%%%%%%

Since the seminal work of Car and Parrinello~\cite{car85} the field of
ab initio molecular dynamics (AIMD) has become impressively
popular as it offers insights into finite-temperature
  systems. Because of the high computational cost
involved in ab initio electronic structure calculations, AIMD is
usually based on density functional theory (DFT)~\cite{kohn65}, which
provides a good compromise between accuracy and computational cost.
However, approximate DFT functionals do not systematically reach
chemical accuracy (1 kcal/mol) and are affected by a series of known
shortcomings~\cite{cohen11}. This can hamper understanding of systems
for which DFT approaches are insufficient to obtain a quantitative,
or even qualitative, standard.

An alternative to DFT is to use correlated wave-function
methods, which are more accurate and systematically improvable, but
involve a higher computational cost.  For applications to molecular
systems the M\o ller-Plesset
perturbation theory~\cite{moller34} (MP2, MP3 etc.) and
coupled-cluster theory~\cite{bartlett07} (CC) have long been
used~\cite{szabo12}. For condensed phase applications,
MP2 has only recently appeared -- using
plane-wave~\cite{marsman09,dixit2017communication},
localized~\cite{pisani08}, and hybrid~\cite{delben2012} basis sets.
Approximations based on coupled-cluster theory have been implemented
in the VASP~\cite{booth13} and PySCF codes~\cite{mcclain17,sun18}.  

A particularly promising correlated approach for condensed phase
applications is the random phase approximation
(RPA)~\cite{bohm53,gell57,langreth75,dobson99,furche01,eshuis11,%
hesselmann11_2,ren12_2}. This
approach to the ground state electronic correlation energy can be seen
both as a DFT approximation~\cite{langreth75,langreth77} or a
CC approximation~\cite{scuseria08}. The RPA also has
analogies with MP theory~\cite{rocca16}.
Unlike CC and MP theories, which
are applied exclusively as post Hartree-Fock methods, the RPA
is most often evaluated from a DFT starting point.  

The success of the RPA
is largely related to its accurate description of weak van der Waals
(vdW) interactions~\cite{dobson99,furche01}, but this approximation is
sometimes less reliable for short range interactions.
Although significantly more expensive than traditional DFT
approximations, the RPA can be performed for systems with up to a few
hundred electrons and a series of applications to condensed phase
systems have been reported in the literature~\cite{delben15,marini06, lebegue10,harl08,lu09,harl10,kaoui16, ren09, schimka10, goltl12}.  
When extra
precision is required, beyond-RPA methods can be used
\cite{furche05,hesselmann11,bates13,olsen12,ren13,lu14,colonna14,rocca16,dixit17,hellgren18}.  

Whether via RPA, MP or CC theories, correlated wave-function methods
hold the promise of a significantly higher level of accuracy, which is
of fundamental importance for design of truly predictive approaches to
assist experimental and technological developments.  However, they are
characterized by an unfavorable scaling, slow basis set
convergence~\cite{marini06, harl08, dixit18}, and slow cell size
convergence~\cite{liao16}, making them expensive.

Despite these difficulties,
high-level methodologies are of vital importance
in many technologically attractive cases
involving e.g. nanoporous materials such as zeolites
\cite{VanSpeybroeck2015,Chibani2016,Bucko2017,Chibani2018,Grajciar2018}
or metal-organic frameworks \cite{Chibani2018,Grajciar2018,Babaei2016,Adil2017}.
These materials are 
intensively investigated for their ability to selectively adsorb
target compounds and this property is used in numerous applications such as 
depollution \cite{VanSpeybroeck2015,Chibani2016,Bucko2017,Chibani2018,%
Grajciar2018,Babaei2016}
or separation of chemicals
\cite{Adil2017,Sholl2016}.
Reliable predictions of adsorption properties of these materials 
can be made only if the electronic structure method used in simulations accurately describes 
all interactions within the system of interest,  
and if the thermal effects are properly accounted for by using a suitable
statistical mechanics method such as  molecular dynamics (MD) or 
Monte Carlo.
Unfortunately, simulations combining high-level
methodologies with MD or Monte Carlo approaches, usually requiring 
ensembles of tens of thousands of configurations, are
  thus far beyond the reach of 
the computational power available today.
In fact, the forces for correlated methods
in the condensed phase have only recently been
implemented~\cite{delben15_2,ramberger17} and employed in MD
simulations~\cite{delben15,bokdam17}.

In this work we present a methodology based on perturbation theory 
that extends the applicability of correlated methods to compute finite
temperature properties at a reasonable computational cost.
Our approach starts from MD
simulations based on computationally inexpensive DFT functionals. Then, a
small number (a few tens) of significant configurations are chosen
from the ensemble of structures generated by MD
to represent the full probability
distribution. Finally, correlated wave-function calculations are
performed on these configurations and the results are used to
reconstruct the corresponding probability distribution and ensemble
averages. Below we will apply
this method to the computation of the finite temperature adsorption
enthalpies of CH$_4$ and CO$_2$ in the
zeolite chabazite, using RPA.

The new approach introduced in this letter
will be referred to as perturbation theory on selected configurations (selPT).
A detailed discussion of perturbation theory in the context of molecular dynamics and free energy calculations can be found in Ref.~\onlinecite{chipot07}.
The main ideas and working equations will be summarized here.

Within the canonical ensemble, the expectation value of a certain
observable $O$ is defined as follows:
$\langle O \rangle_{H} =
\int O(\vq,\vp) \text{exp}\left\{- \beta H(\vq,\vp) \right \}  d\vq\,d\vp
/Z$, for
$Z=\int \text{exp}\left \{ -\beta H(\vq,\vp) \right \}  d\vq\,d\vp$.
Here $\beta=1/k_B\,T$ and $\vq$ and $\vp$ denote nuclear positions
and momenta, respectively. 
In this work the notation $\langle \ldots \rangle_{H}$ represents the
canonical ensemble average corresponding to the classical Hamiltonian
$H(\vq,\vp)=T(\vp)+V(\vq)$ with $T(\vp)$ and $V(\vq)$ being the
nuclear kinetic and potential energies, respectively.

Exploring all $\vp$ and $\vq$ is generally impossible.
An alternative is to assume the ergodic hypothesis. Then, the
ensemble average can more usefully be expressed as
\begin{eqnarray}\label{ergo}
\langle O \rangle_{H}
 = \lim_{\tau\to\infty}\frac{1}{\tau}
 \int_0^\tau O(\mathbf{q(t)},\mathbf{p(t)}) \,dt,
\end{eqnarray}
a temporal average over an AIMD trajectory
$\dot{\vq}=\nabla_{\vp}H$, $\dot{\vp}=-\nabla_{\vq}H$. For infinite
time, $\langle O \rangle_{H}$ is independent of trajectory, since
the system explores all configurations.
However, for finite time, the choice of trajectory is crucial.

Ideally, we could use any
method to obtain long AIMD trajectories that ensure a sufficient
level of statistical precision. AIMD simulations
are time-consuming, however, and thus long trajectories
require a numerically efficient model electronic energy
Hamiltonian $\hat{H}_{e}$, with potential
\begin{align}
V(\vq)=&\langle \Psi(\vq) |\hat{H}_{e}| \Psi(\vq)\rangle
+ U_{II}(\vq),
\label{vq_abinitio}
\end{align}
at nuclear configuration $\vq$ and repulsive nuclear energy $U_{II}$.
Unfortunately, this leaves us with a compromise between
accuracy (quality of electronic energies) and efficiency
(length of trajectories), for a given precision.

In such cases, perturbation theory can be used to transform
from an efficient/inaccurate Hamiltonian $H=T+V$ (e.g. PBE+D2) to
an expensive/accurate, Hamiltonian $H'$ (e.g. RPA), via the
perturbation,
\begin{align}
  H'(\vq,\vp)=&H - V + V'
  \equiv H(\vq,\vp)+\Delta V(\vq).
\end{align}
The trick is to write thermal averages of potentials as,
\begin{align}
  \langle V \rangle_H =& \int E dE P_H^{V}(E),
  ~~
  P_H^{V}(E) \equiv \langle \delta(V-E) \rangle_H,
  \label{eq_distribMoment}
\end{align}
where $P_H^{V}(E)$ describes the energy $E$ distribution of
some potential energy $V$ of interest. The subscript $H$ indicates
that it is calculated in the trajectory described by Hamiltonian $H$.
Then, one uses the perturbation $\Delta V(\vq)$ to obtain
$\langle V' \rangle_{H'}=\int EdE P_{H'}^{V'}(E)$ from
\begin{align}\label{distribution2}
  P_{H'}^{V'}(E)=&
  \frac{\iexpect{ \delta(V'-E)\exp\{-\beta\Delta V\}}_H}%
  {\iexpect{ \exp\{-\beta \Delta V\} }_H}\;.
\end{align}
$P_{H'}^{V'}$ then gives properties of the trajectory $H'$, but is
calculated on the trajectory $H$.

This lets us define properties of the system driven by the 
primed Hamiltonian (e.g. RPA)
in terms of trajectories (Eq.~\ref{ergo}) on the 
system described by the original Hamiltonian (e.g. PBE+D2).
Typical calculations employing perturbative
approaches based on this formula %, e.g. as available in the literature,
involve large (albeit reduced)
numbers of configurations computed in a given MD or
Monte Carlo simulation, which are implausible for the expensive
methods we seek to use. In this work, we thus introduce an innovation
that lets the perturbation be evaluated with very few additional
calculations, but without significant loss of precision.
We thereby break the compromise between speed and accuracy
that hampers the application of AIMD to difficult problems,
such as the ones presented below.

%%%%%%%%%%%%%%%%%%%%%%%%%%%%%%%%%%%%%%%%%%%%%%%%%%%%%%%
\hide{
\begin{figure}[h!]
\begin{center}%
\includegraphics[width=0.4\columnwidth]{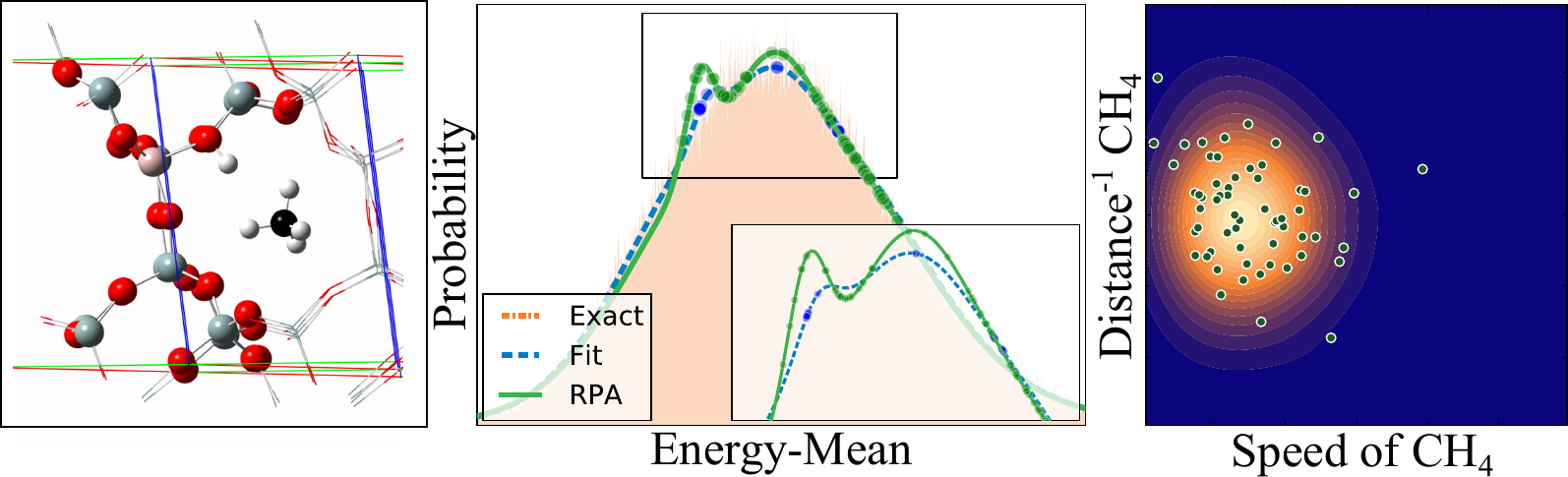}%
\end{center}
\caption{The model of CH$_4$ adsorbed on protonated chabazite
  studied in this work. Color code: Al in pink, O in red, Si in blue, C in black,
and H in white.}\label{fig_chab}
\end{figure}
}
\begin{figure}[t!b!]
  \includegraphics[width=\columnwidth]{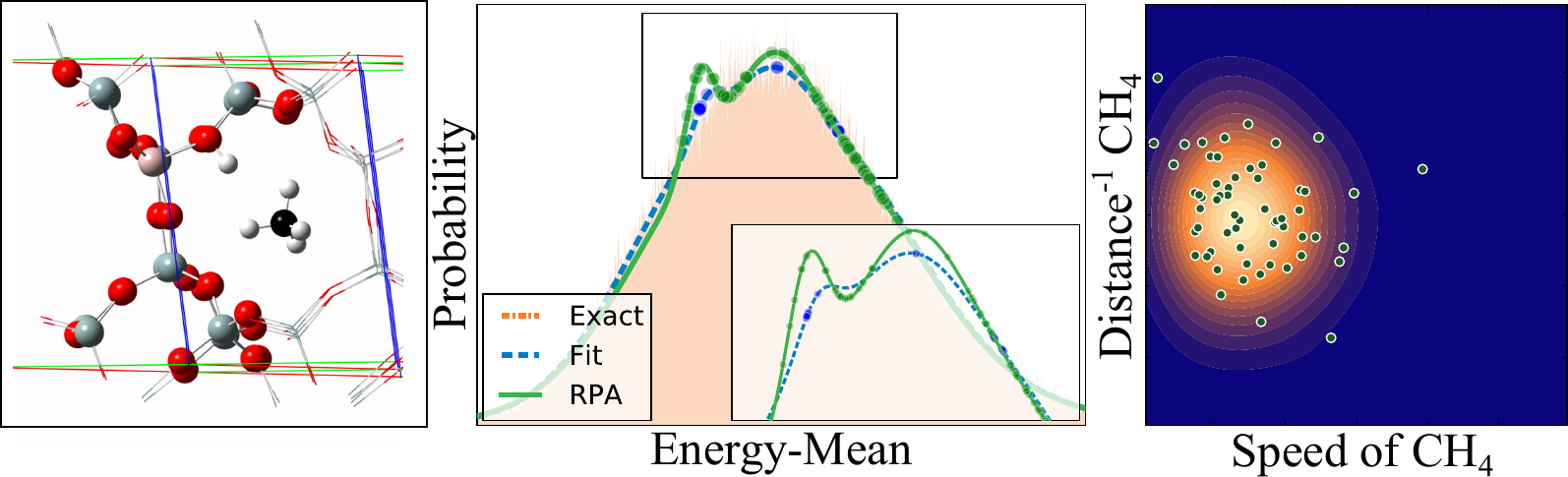}%
  \caption{
    Left: The model of CH$_4$ adsorbed on protonated chabazite studied in
    this work. Color code: Al in pink, O in red, Si in blue, C in black,
    and H in white.  Middle: Probability distributions of the energy for
    CH$_4$ in protonated chabazite at T=300K. Compares the exact
      distribution (2000 bins, shaded orange) from a PBE+D2 MD run, with
      its selPT counterpart obtained using 60 configurations (SET2@PBE+D2,
      blue dashed line), and perturbed onto RPA (green solid line).  All
      distributions are shown relative to their mean. Right: Phase space
    of the PBE+D2 trajectory projected on CH$_4$ speed and a metric
    representing the inverse distance of the CH$_4$ from the
    chabazite. The 60 selPT states are shown as circles in both
      panels.}\label{fig_all}
\end{figure}
%%%%%%%%%%%%%%%%%%%%%%%%%%%%%%%%%%%%%%%%%%%%%%%%%%%%%%%

To understand the challenges involved in the direct application of
perturbation theory, we now present a specific numerical example that
will also be used later to discuss our new methodological
developments. Let us consider the finite temperature adsorption of
methane on protonated chabazite (later, results will also be
presented for the adsorption of carbon dioxide).  
The model used for this
system, which involves 200 valence electrons per unit cell, is shown in
Fig.~\ref{fig_all}
and further discussed in Supplemental Material.  Using MD in the NVT ensemble, the adsorption
enthalpy can be computed as
\begin{equation}\label{ads}
  \Delta_{\ads}H(T)= \iexpect{V_{S+A}}_H - \left [ \iexpect{V_{S}}_H +
    \iexpect{V_A}_H \right ]- k_BT,
\end{equation}
where subscripts $S$ and $A$ denote the clean substrate (protonated
chabazite), and the adsorbate (methane), respectively, while $S+A$
indicates the system of substrate interacting with adsorbate.  In
order to evaluate the ensemble averages in Eq.~\ref{ads}, MD
simulations at 300~K have been performed with the VASP
code~\cite{kresse93} using two different approximations, namely PBE
and PBE+D2 (a vdW corrected version of the PBE
functional~\cite{grimme06}). Computational details are provided in
Supplemental Material.  These two MD simulations will be denoted as
MD@PBE and MD@PBE+D2, respectively.

In Tab.~\ref{table1} the PBE (PBE+D2) enthalpy of adsorption
corresponding to MD@PBE (MD@PBE+D2) is an ``exact'' estimate,
namely an estimate
where the averaged potential and the Hamiltonian used for
the sampling are at the same level of theory; the PBE (PBE+D2) value
evaluated from MD@PBE+D2 (MD@PBE)  is based on the perturbative
approach in Eq.~\ref{distribution2}.
Both PBE+D2 and PBE values evaluated perturbatively from MD@PBE and MD@PBE+D2, respectively,
are close to their ``exact'' values, demonstrating the reliability
of our approach. The PBE+D2 perturbative estimate is slightly worse, showing the importance
of starting from an approach that correctly samples the key
adsorption region of phase-space.

From a practical point of view, this brute force
application of perturbation theory involves the reevaluation of the
energy for thousands of configurations (specifically, the
reevaluation of the electronic part of Eq.~\ref{vq_abinitio}).
If, starting from MD@PBE or MD@PBE+D2, we applied this
perturbative procedure to compute the RPA enthalpy of adsorption, the
corresponding computational time would amount to several millions of
CPU hours in our supercomputer center. While technically feasible,
such high demands are of limited practical interest.
  
We now proceed to show that it is possible to reduce the number of
times the electronic part of Eq.~\ref{vq_abinitio} is recomputed to a
very small fraction of the total number of configurations. This is
achieved by finding a semi-analytic model of $P_V^H(E)$, and using it
to build an approximate model of $P_{V'}^{H'}(E)$
by assuming that the distributions generated by
$H$ and $H'$ (e.g., $H_{\PBED}$ and $H_{\RPA}$, as used here)
are similar.
We employ the following procedure to approximately reconstruct the
correct high-level energy distribution $P_{H'}(V')$ and
corresponding average values:%
%%%%%%%%%%%%%%%%%%%%%%%%%%%%%%%%%%%%%%%%%%%%%%%%%%
\newcommand{\Step}[1]{\vspace{1mm}\\{\bf Step #1}:~}
%%%%%%%%%%%%%%%%%%%%%%%%%%%%%%%%%%%%%%%%%%%%%%%%%%
%\begin{enumerate}
%\item
\Step{1}
MD driven by the Hamiltonian $H(\vq,\vp)$ is performed to generate
the $P_H^V(E)$ distribution (eq.~\ref{eq_distribMoment}).   
%\item
\Step{2}
$P_H^V(E)$ is approximated by a set of $N_G$ Gaussian
functions with $3N_G$ fitting parameters 
$\left \{h_i, w_i,\epsilon_i \right \}$
\begin{equation}\label{eq_h1DistribApprox}
  \Pt_H^V(E) =  \frac{1}{M} \sum_{i=1}^{N_G} h_i\, 
  \exp\bigg\{ -\frac{\left (E - \epsilon_i \right )^2}{2w_i^2}
  \bigg\},
\end{equation}
with normalization constant $M=\sum_{i=1}^{N_G} \sqrt{2\pi}h_i
w_i$. Parameters are selected to represent key statistical
features of the distribution.
The tilde introduced in Eq.~\ref{eq_h1DistribApprox} emphasizes the 
approximate character of the probability distribution.
It is crucial here to forbid any single or a limited number of
Gaussians to dominate; for this reason the contribution of each single
Gaussian is limited to a certain threshold.
We note that this representation of potential energy density is related to 
the kernel density estimation method of probability density functions~\cite{Parzen:1962}. 
Technical details on the fitting with Gaussian functions are further discussed in Supplemental Material.
%\item
\Step{3}
  $N_G$ configurations with positions $\vq_i$ are selected from
  the ensemble of step~1 to satisfy
  \begin{equation}~\label{condition1}
      V(\vq_i)\in [\epsilon_i-\Delta, \epsilon_i+\Delta]\;,
  \end{equation}
where we use
    $\Delta=0.02$~kcal/mol. Since
 several time-uncorrelated configurations can satisfy
 Eq.~\ref{condition1}, the specific
configuration $\vq_i$ is
randomly chosen with uniform probability. This is a reasonable procedure since configurations
with the same energy ($\epsilon_i$) have equal probability to be sampled.
For the selected configurations $\vq_i$
a different (higher) level of theory is used to evaluate the energies $V'(\vq_i)$
to obtain $\Delta V_i \equiv \Delta V(\vq_i) = V'(\vq_i)-V(\vq_i)$.
%\item
\Step{4}
Next, we calculate
  \begin{align}\label{eq_h2Shifts}
    {{\epsilon}_i}'=&\epsilon_i + \Delta V_i\;,
    &
    h'_i =& h_i\exp\{-\beta\Delta V_i\}\;,
    \end{align}
  and use them to obtain:
  \begin{equation}\label{eq_h2DistribApprox}
    \Pt_{H'}^{V'}(E) = \frac{1}{M'} \sum_{i=1}^{N_G} h'_i\,
    \exp\bigg\{
    -\frac{(E - {{\epsilon}_i}' )^2}{2w_i^2}
    \bigg\}\,,
  \end{equation}
  where $M'=\sum_{i=1}^{N_G} \sqrt{2\pi}h'_i w_i$.
  This equation follows from Eqs.~\ref{distribution2} and \ref{eq_h1DistribApprox}, with
  reweighting factor $\exp\{ - \beta \Delta V \}$. 
%\item 
\Step{5}
Finally, we determine the estimate of $\iexpect{{V'}}_{H'}$  
by combining eq.~\ref{eq_h2DistribApprox} with eq.~\ref{eq_distribMoment}.
The accuracy of this ensemble average can be evaluated by computing the standard error 
according to Eq. S17 (see Supplemental Material). If a higher level of accuracy is required 
the original set of selected configurations can be extended to
decrease the statistical error of the final result, as done later.
%\end{enumerate}

We label the five step procedure described above by the acronym
``selPT'', which stands for \emph{perturbation theory on selected
  configurations}.  In order for this approach to be useful,
statistical convergence should be achieved for $N_G\ll N_{\text{MD}}$,
where $N_{\text{MD}}$ denotes the total number of MD steps. This is
not an unreasonable expectation, as the configurations produced within
a certain MD simulation are not completely independent due to the
finite correlation time.  Below, we show that even $N_G=60$
configurations can be
  sufficient for a reasonable estimate and $N_G=180$ can provide
  highly accurate results -- both
  are vast improvements on the $N_{\text{MD}}\approx 190,000$ runs
required to recompute
a full MD trajectory. Such a dramatic reduction is
possible because $\Pt_{H'}^{V'}(E)$ is a smooth, representative
function that captures the important properties of the true
distribution $P_{H'}^{V'}(E)$.

%%%%%%%%%%%%%%%%%%%%%%%%%%%%%%%%%%%%%%%%%%%%%%%%%%%%%%%
\hide{
\begin{figure}[t!b!]
\includegraphics[width=\columnwidth]{Fig2}%
\caption{
Left: Probability distributions of the energy for CH$_4$ in
protonated chabazite at T=300K. Compares the exact distribution
  (2000 bins, shaded orange) from a PBE+D2 MD run, with its
  selPT counterpart obtained using 60 configurations (SET2@PBE+D2,
  blue dashed line), and perturbed onto RPA (green solid line).
  All distributions are shown relative to their mean.
Right: Phase space of the PBE+D2 trajectory
projected on CH$_4$ speed and a metric representing the inverse
distance of the CH$_4$ from the chabazite. The 60 selPT states are
shown as circles in both panels. Inset text reports calculation times
for key steps.}\label{fig_distribApprox}
\end{figure}
}
%%%%%%%%%%%%%%%%%%%%%%%%%%%%%%%%%%%%%%%%%%%%%%%%%%%%%%%

To establish the accuracy and the efficiency of the selPT procedure 
we come back to the example of CH$_4$ adsorbed on chabazite. 
Before computing RPA adsorption energies, the approach is validated
using PBE and PBE+D2: First, the ``exact'' distribution $P_{H_{\PBED}}^{V_{\PBED}}$
generated by the previously described MD@PBE+D2 simulation is fitted
by 60 Gaussian-shaped functions (eq.~\ref{eq_h1DistribApprox});
second, 60 configurations $\vq_i$ are chosen in order to satisfy the
condition in Eq.~\ref{condition1}; finally,
Eq.~\ref{eq_h2DistribApprox} is used to evaluate $\Pt_{H_{\PBE}}^{V_{\PBE}}$
and the corresponding ensemble averages. The procedure is repeated
three times to generate three uncorrelated sets containing 60 configurations each, that
will be called SET1@PBE+D2, SET2@PBE+D2, and SET3@PBE+D2. This is helpful to study the behavior of the selPT method
as a function of the number of configurations. Indeed, the adsorption enthalpies have been computed using the three sets with 60 configurations, 
from the three sets with 120 configurations obtained by merging pairs of the original sets (SET1+2@PBE+D2, SET2+3@PBE+D2, and SET1+3@PBE+D2), and with a set
that includes all the configurations (SET1+2+3@PBE+D2). 
Since the enthalpy of adsorption Eq.~\ref{ads} requires ensemble averages for the three different
systems (S, A, S+A), the selPT procedure is applied independently on each one of them. 
%\TB{}{
%As discussed in Sec. SV of the Supplemental Material, 
%the selPT calculations of the RPA energy based on 
%the configuration space sampling via MD@PBE 
%are less efficient compared to those employing MD@PBE+D2.
%In our target application presented below, 
%we therefore use only the MD@PBE+D2 as the sampling method 
%(the results based on the MD@PBE sampling are provided in the Supplemental Material).}

Fig.~\ref{fig_all}
demonstrates the fitting quality.
The middle
panel demonstrates the ability of selPT to reproduce
key features of the sampling data, and to perturb them using a more
accurate method, by comparing $P_H^V$ from the full
MD@PBE+D2 run, with its parametrized distribution
$\tilde{P}_H^V$ and RPA counterpart $\tilde{P}_{H'}^{V'}$
approximated from Eq.~\ref{eq_h2DistribApprox}
using SET2@PBE+D2.
The right panel shows a phase space diagram
for the full MD run, and the 60 samples selected by selPT, reported in
terms of variables representing the speed of CH$_4$ and its distance
from the surface of the chabazite. Note that selPT samples the
probability distribution accurately, by design; as shown in the right panel of
Fig.~\ref{fig_all}
a representative 
part of phase space is also covered for the system under consideration.

%%%%%%%%%%%%%%%%%%%%%% Table 1 %%%%%%%%%%%%%%%%%%%%%%%%
\begin{table*} %[b!]
\caption{\label{table1}
  Enthalpy of adsorption $\Delta_{\ads}H(T)$ (in kcal/mol) of CH$_4$
  in protonated chabazite at 300 K evaluated at the PBE+D2, PBE, and
  RPA levels of theory. Results are calculated using full MD
  trajectories (MD@PBE and MD@PBE+D2) applying the selPT method to a
  set with $N_G=60$ configurations (SET1@PBE+D2),
  $N_G=120$
  (SET2+3@PBE+D2), and 
  $N_G=180$
  (SET1+2+3@PBE+D2).  Experimental results at room temperature give
  $\Delta_{\ads}H(T)=-4.06$~kcal/mol \cite{piccini15, luo16}.}
\begin{ruledtabular}\begin{tabular}{c|cccccc}
%    & \multicolumn{6}{c}{Trajectory ($H$)} \\
  \multirow{2}{*}{\backslashbox{\ $V$ \ }{\ $H$\ }}  & MD & MD & SET1 & SET2+3 & SET1+2+3 \\
    & @PBE & @PBE+D2 &
    @PBE+D2  & @PBE+D2 & @PBE+D2
    \\\hline
		PBE+D2   & $-5.72\pm0.48$ & $-6.09\pm0.18$ & - & - & - 
    \\
    PBE      & $-1.13\pm0.18$ & $-1.11\pm0.25$ & $-1.31\pm0.58$ & $-1.84\pm0.53$ & $-1.73\pm0.43$ 
    \\

    RPA      &    -- &    -- & $-4.20\pm0.62$ & $-4.45\pm0.48$ & $-4.38\pm0.39$ & 
    \\
\end{tabular}\end{ruledtabular}
\end{table*}

\hide{
\begin{table} %[b!]
\caption{\label{table1} 
  Enthalpy of adsorption $\Delta_{\ads}H(T)$ (in kcal/mol) of CH$_4$
  in protonated chabazite at 300 K evaluated at the PBE, PBE+D2, and
  RPA levels of theory. Results are calculated using full MD
  trajectories (MD@PBE and MD@PBE+D2) applying the selPT method with a
  set with $N_G=60$ configurations (SET1@PBE+D2),
  $N_G=120$
  (SET2+3@PBE+D2), and 
  $N_G=180$
  (SET1+2+3@PBE+D2).  Experimental results at room temperature give
  $\Delta_{\ads}H(T)=-4.06$~kcal/mol \cite{piccini15, luo16}.}
\begin{ruledtabular}\begin{tabular}{c|cccccc}
%    & \multicolumn{6}{c}{Trajectory ($H$)} \\
  \multirow{2}{*}{\backslashbox{\ $V$ \ }{\ $H$\ }}  & MD & MD &
  $N_G=60$ & $N_G=120$ & $N_G=180$ \\
    & @PBE & @PBE+D2 &
    @+D2  & @+D2 & @+D2
    \\\hline
    PBE      & $-1.13$ & $-1.08$ & $-1.20$ & $-0.00$ & $-0.00$ 
    \\
    Err & $\pm0.00$ & $\pm0.00$ & $\pm0.XX$ & $\pm0.XX$ & $\pm0.XX$
    \\
    +D2   & $-5.72$ & $-6.09$ & $-5.72$ & $-0.00$ & $-0.00$ 
    \\
    Err & $\pm0.00$ & $\pm0.00$ & $\pm0.XX$ & $\pm0.XX$ & $\pm0.XX$
    \\
    RPA      &    -- &    -- & $-4.20$ & $-4.45$ & $-4.38$ & 
    \\
    Err & -- & -- & $\pm0.62$ & $\pm0.48$ & $\pm0.39$
    \\
\end{tabular}\end{ruledtabular}
\end{table}
}

%%%%%%%%%%%%%%%%%%%%%%%%%%%%%%%%%%%%%%%%%%%%%%%%%%%%%%%

The row corresponding to PBE in Table~\ref{table1} shows the practical reliability
of selPT in computing enthalpies
of adsorption $\Delta_{\ads} H(T)$.
Specifically, the similarity of the energies across all
calculations highlights the success of the approach. If our
trajectories sampled an overly-limited set of configurations 
(whether through a too-short trajectory or too-inaccurate functional),
or if our selections sets were too poor, the numbers would differ
significantly across columns. The consistency of the values justifies
the length and reasonableness of our initial AIMD calculations, and
the accuracy of the selection process.

The bottom row of Table~\ref{table1} shows RPA enthalpies of
adsorption, which would be basically impossible to obtain using a full
AIMD treatment due to the inherent cost of RPA (full results considering all the possible combinations
of sets are provided in Tables S9 and S10 of Supplemental Material). Numerical results are
based on the RPA implementation in the VASP code~\cite{harl08} and
computational parameters are provided in the Supplemental Material.
The statistically most accurate selPT estimate of the RPA enthalpy of adsorption is obtained 
from the largest set SET1+2+3@PBE+D2 ($N_G=180$) and provides a value of $-$4.38 $\pm$ 0.39 kcal/mol
in excellent agreement with the experimental value of $-4.06$~kcal/mol~\cite{piccini15,luo16}. 
From Table~\ref{table1} (and Tables S9-S10) it can be noticed that the errors present 
a clear trend to decrease as a function of the number of configurations included in the selPT set.
In Table S9 it is shown that RPA results
from as few as 60 samples are all within 0.92~kcal/mol of one
another, and within
0.83~kcal/mol of experiment. This is already a considerable improvement with respect to the
PBE and PBE+D2 estimates, which present deviations of
at least 2 kcal/mol well beyond the chemical accuracy.

Key for this high level of accuracy is that the selPT calculations are
based on a reasonable PBE+D2 trajectory which, unlike PBE, samples
phase space similarly to RPA. This ensures that the perturbation from
PBE+D2 to RPA is correctly biased to the weak vdW interaction between
the zeolite and molecule. This intuitive point is further discussed in
a quantitative way in Supplemental Material where RPA results from a
PBE starting point are presented (Secs. SIII and SIV) and
it is shown
how PBE leads to significantly different sampling of the
configurational space with respect to PBE+D2 and RPA (Sec. SV).

To further establish the accuracy and reliability of the selPT approach, 
a second system is discussed, CO$_2$ adsorbed in protonated chabazite. 
Computational details are analogous to those of CH$_4$
in chabazite. 
The adsorption enthalpy of CO$_2$ is sizeably larger than that of CH$_4$, with an experimental value of -8.41 kcal/mol
at room temperature~\cite{pham12}. While the PBE+D2 result deviates from this value by 1.3 kcal/mol, our most accurate estimate from the largest set SET1+2+3@PBE+D2  
provides an highly accurate enthalpy of adsorption of $-$8.74 $\pm$ 0.39 kcal/mol.
Detailed results for all the test sets are presented in Table S9 ($N_G=60$) and S10 ($N_G=120$) in Supplemental Material.
The errors are analogous to those for adsorbed CH$_4$;
by considering that the adsorption enthalpy of CO$_2$ is about double than that of CH$_4$,
the relative error in this case decreases much more rapidly as a function of the set size and $N_G=60$ provides already an accuracy within $\pm10\%$. 

From a practical point of view, the selPT approach has been
instrumental to achieve the computation of RPA enthalpies of
adsorption. By decreasing the number of configurations to 60, the RPA
results are obtained in a CPU time that is only about 3-4 times the
CPU time involved in the full MD@PBE calculation, or around ten
times for the more accurate simulation with 180 configurations.

In conclusion, we introduced the selPT method to compute 
ensemble
averages of energies and, more generally, energy distributions for
correlated wave-function methods.  As proof of principle, this
methodology was applied to compute the finite-temperature adsorption
energy of CH$_4$ and CO$_2$ in protonated chabazite at
the RPA level of theory,
where it successfully reproduced known experimental values within chemical accuracy. 
selPT is
not limited to the RPA but could be interfaced with other
sophisticated approximations (e.g. MP2, double-hybrids or
coupled-cluster theory), as long as computations on few tens of
configurations can be afforded.

Beyond providing an accurate and efficient methodology to address
challenging problems in physics and material science, selPT could be
also used to develop finite-temperature test sets from high level
theories, to test and develop new DFT functionals and
  semi-classical models.

\begin{acknowledgments}
This work was supported by Agence Nationale de la Recherche under grant number ANR-15-CE29-0003-01.
T.B. is grateful to University of Lorraine for invited professorships
during the academic year 2015-16 and he
acknowledges support from Project
643 No. APVV-15-0105.
D.R., T.G. and S.L. acknowledge support from the French PIA project “Lorraine Universit\'e d'Excellence.
The results of this research have been achieved using GENCI-CCRT/CINES computational resources under grants A0030805106 and A0040910433, the DECI resource ARCHER based in the United-Kingdom with support from the PRACE aisbl, and the Computing Center
of the
Slovak Academy of Sciences acquired in projects ITMS 26230120002 and 26210120002 supported by the
Research and Development Operational Program funded by the ERDF.
\end{acknowledgments}

\bibliographystyle{apsrev}
%\bibliography{rpa_ref,mof_zeo}

\end{document}